# The 3D virtual environment online for real shopping

Nahla Khalil

Computer science and Information Technology Department

Computer Solution Center

Toronto, Canada

khalilnahla@yahoo.com

*Abstract:* **The development of information technology and Internet has led to rapidly progressed in e-commerce and online shopping, due to the convenience that they provide consumers. E-commerce and online shopping are still not able to fully replace onsite shopping. In contrast, conventional online shopping websites often cannot provide enough information about a product for the customer to make an informed decision before checkout. 3D virtual shopping environment show great potential for enhancing e-commerce systems and provide customers information about a product and real shopping environment. This paper presents a new type of e-commerce system, which obviously brings virtual environment online with an active 3D model that allows consumers to access products into real physical environments for user interaction. Such system with easy process can helps customers make better purchasing decisions that allows users to manipulate 3D virtual models online. The stores participate in the 3D virtual mall by communicating with a mall management. The 3D virtual mall allows shoppers to perform actions across multiple stores simultaneously such as viewing product availability. The mall management can authenticate clients on all stores participating in the 3D virtual mall while only requiring clients to provide authentication information once. 3D virtual shopping online mall convenient and easy process allow consumers directly buy goods or services from a seller in real-time, without an intermediary service, over the Internet. The virtual mall with an active 3D model is implemented by using 3D Language (VRML) and asp.net as the script language for shopping online pages**

*Keyword:* Online Shopping; 3D; Virtual Reality; retailing; E-Commerce; virtual environment

1. INTRODUCTION

The tremendous development of technology take place since the invention of online shopping by Michael Aldrich [1,2] in the UK followed the Worldwide Web **[** 3,4,5] wrote by Tim Berners-Lee and gave the first browser to view the web which changed most of things; a whole new revolution started. Later, Netscape **[ 6]** released Navigator browser, and introduced Secure Sockets Layer (SSL) **[7,8]** encryption for secure transaction. Then transactions and business started happening over the web urging every company to have a ".com". To address security issues Netscape 1.0 introduced SSL encryption, for secure transaction. When money was involved in doing business over the internet "PayPal" was launched, it provides facility for online payment. E-commerce websites, such as Amazon.com, Dell.com, Pizza Hut.com and eBay.com, started selling their products online, also Search engine such as Yahoo and Google, Facebook.com, and others. All these Companies are created 2D Two-Dimensional environments website. These business activities have redefined products, distribution channels, and industries [9]. Quickly, Online shopping websites become very attractive way to use by customers due to their advantages, such as a convenient price, easy, a fast shipping, price comparison, obtaining customer reviews, and more recently social shopping [10], especially in developed countries. Moreover, these online shopping sites turn into very popular amongst customers because it considered entertaining and time consuming to the customers. Developing online shopping sites and added more facilities, such us transforming these 2D sites in to three-dimensional (3D) through the additional use of virtual reality. Theses development has led to a gradually use of the 3D virtual reality in different application, and attract customers in different fields such as Games, Movies and etc. The information development has greatly modified business and commerce, education, security, and social interaction [11]. Additionally, the 3D virtual shopping environment gathers several online stores under one roof to create an environment where the customers can traverse from one store to another easily. Creating a virtual shopping mall helps to improve the customer's satisfaction through transforming website to 3D and using virtual reality, the computer usage evolved to the next level of technology, that is considered helpful and can create a more efficient, less costly, and higher-quality service-delivery environment for the users [11]. Using 3D virtual environments replicating the look touch and feel experience of shopping that allows consumers to have an in store experience online. These environments give merchants the chance to close the gap between online retail and traditional real shopping. Ultimate 3D virtual shopping environments, offering consumers something that regular online shopping cannot offer. 3D virtual shopping environments offer a shopping solution that mirrors the real world shopping experience. Shopping is more than just buying; it's the excitement, the noise, the look, and the colors. Simulate this experience online by bringing everything customers experience in the real world to the comfort of their home. It combines physical shopping and online shopping together from the ease and comfort of consumers' homes. In this work created interactive 3D computer world that can explore the feeling and reality of in store shopping environment with both mentally and physically.

## 2. LITERATURE REVIEW

The Internet has changed many aspects of our life and becomes one of the requirements for our daily life, for being a transformation channel to the world of accessing to information. So, the user can access information that are available on the Internet easily, and carry out different activities through the use of Internet services like e-business, e-commerce, e-learning, e-governance activities and etc. The possibilities of sharing and accessing information that is available on the Internet are mostly contributing to the users' social life and daily activities. Online services have become a viable alternative for different actions like: online shopping, educations, meetings, and organizing businesses etc. General 2D websites, containing width and height coordinates, currently used with most online shopping facilities are considered interesting to the users and provide several market activities and actions of selling, buying and advertising over the Internet. Due to the development of information technology (IT) and Internet, Business activities could expand 2D web environments to be used in 3D virtual world environments [14]. Where 3D virtual worlds have the potential to revolutionize business and bring significant implications and activates to business including opportunities for co-creation and enhancing customers' perceptions and value of a brand [12]. The idea of the virtual world environments technology can be used to create environments that are closer to the field and easily implement in many important applications.

Such system can be used for entertainment, simulation, and education. Virtual Reality (VR) is going to change the way we express ourselves, communicate with each other and experience the world by creating an environment to be explored. With highly advanced devices in development today, such as Oculus Rift and HoloLens, virtual reality and holography is shaping up to be the next generation of 3D technology [29], which offers us more immersive virtual experiences than anything we've ever seen. Everything from video games to live music festivals and VR movies, these new technologies are already changing the world of entertainment in many ways.

3D, computer-generated environment that appears similar to our 'real' world, often massively multi-user can connect to a developed supply online entertainment and social networking for users [13]. In chatting, people can meet and chat with others from all over the globe. Virtual chat site is filled with thousands of fun and interesting people to chat with. Generally, each person has own avatar. An avatar is a 3D representation of each person that other people can see when they are chatting. Avatars even have the ability to run, jump, fly, dance, and express a whole host of emotions and actions [22]. 3D virtual technique offers a rich environment for customers that help to improve the user interface by interacting with each other, and increasing the user motivations to navigate inside realistic looking environments [16]. 3D virtual environments are usually developed for gaming, recreation and entertainment [15]. However, the use of 3D game technologies for the purpose of developing affordable, easy to use and pleasing virtual environments [21]. In addition, 3D gaming has emerged as one of the fastest growing technology-development industries from marathon trivia games to league, bowling nites, there are many sites to provide dozens of creative gaming environments such as soccer, hockey, bowling, chess, checkers, bingo, etc.

Edutainment is the blending of education and entertainment; it is about engaging, enjoyable experiences providing a learning value [20]. Education and technology are interconnected that offer unique learning opportunities. Today millennials feel pretty comfortable with online education, doing research on the Internet, resorting to instructional videos on YouTube and distance learning powered by video technology. Obviously, a 3D virtual world opens the door to a new way of learning. Establishing realistic environments provides a powerful set of learning oriented tools, these platforms allow for the implementation of sophisticated instructional models within a framework with richer information and cooperation [16]. In general, virtual worlds provide rich possibilities for social and behavioral research, including a variety of testing capabilities [19]. Huang et al, Uses the particular VR features in education based 3D technology and discusses learning approaches [17]. Virtual world environment are used now for teaching purposes and it includes tutorials and online lectures. There are many examples of universities and other organizations [27, 28] that are now using virtual worlds for educational and teaching purposes.

Virtual reality is also used for training purposes in the military [23], which includes flight and battlefield simulations, medical training under battlefield conditions, virtual boot camp and more. In the field of Madison, Using virtual reality (VR), for actual diagnostic images of a patient could be used to create a 3D model of the patient [24]. The potential use of these "scientific imaging" models in medical diagnosis, treatment planning, and education will revolutionize the field [25]. The benefits of determining the location of tumors, the placement of surgical incisions, or practicing difficult surgical procedures ahead of time are of inestimable value [26].

The 3D technology provides customers with the ability of viewing and manipulating physical objects using 3D technology so that they can view the items under their favorable environment. At the same time, the agent technology is applied to greatly enrich the customer's shopping process by simulating the body language of the customer's symbol such as their pose and the controller's response in the virtual environment [18].

This paper presents a new type of e-commerce system, which obviously brings virtual environment with an active 3D model. This system implemented by using 3D Language (VRML) to allow consumers to access into real physical environments to enrich shopping experience and user interaction

### 3. RESEARCH METHODOLOGY

This project intend to implement a virtual environment online shopping in 3D model, since all of its floors and shops will be represented in that model, also there will be an automatic camera walk through the mall as if someone do the shopping. That model will be uploaded to the internet through a website, there will be a menu for all the shops categories such as: Women's wear, kids, men's wear … etc, at that menu once any shop clicked, the camera move us to see it.

When the shop door clicked, a webpage opened to see what is there in the shop, customers may pick something to buy and pay really through the credit card online. Sure shipping and other fees will be added and all customer information will be recorded, also there will be a space for customer's recommendations for any item to let other customers read about it and benefit from it.

*3-1 System Requirement*

- VRML ( **Virtual Reality Modeling Language** ) for the 3D model
- HTML for the website
- ASP.net for the shopping and payment pages
- SQL Server for the data base

*3-2. System Analysis:*

*1) The System's Context*

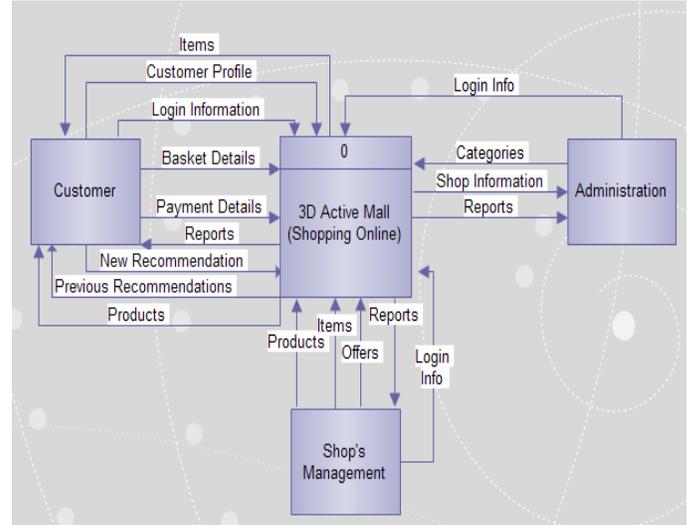

Fig. 1: Context Diagram

The context diagram shows the main entities of the system with all coming and outing data flows, also shows the complete shopping online project include products, items, shops, payment, customers … and more.

**The System's Context** includes three main elements:
A. *Administrator:*

Sending Data:

The administrator login to the system, then send "Outgoing" Data Flow :
- Such us categories of the shops: wear, clothes, babies, cosmetics … etc.

Receiving data:

- Reports about shops and categories.
- Shops information

B. *Shop's Management:*

Sending Data:

- Items information: women clothes, men wear, perfumes…etc.
- Offers Information: sales, buy 1 get 1 free …etc.
- Products Information: That mean certain product, shirt, dress …etc.

Receiving Data:

- Reports about all the items and products in the store, customers, payment movements for certain period of time.

### C. Customer:

Anyone could register in the website and buy anything.

Sending Data:

- Profile information and bank card information.
- Login information (username, password)
- Basket Details: what he wants to buy and the quantities.
- Recommendation about certain item or shop.

Receiving Data:

- Reports about his movements, and payments.
- All products and items details.
- Recommendations from other people.

### 2) Entity Relationship

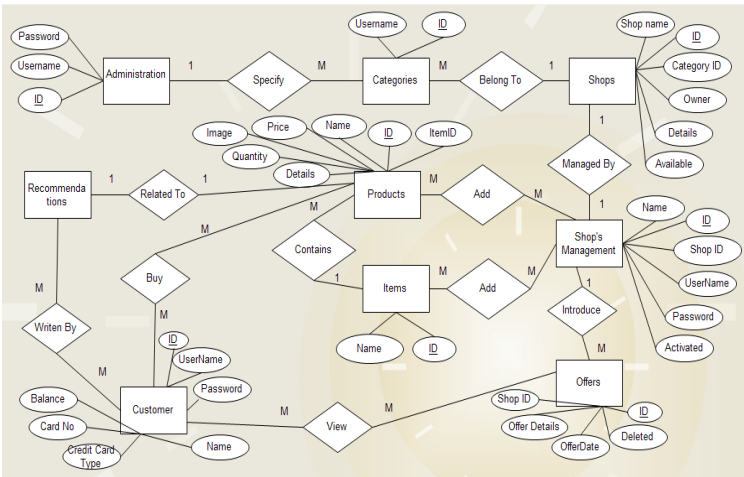

Fig. 2: ER – Diagram

The ER – Diagram reflect the entity relationship between entities: start with administrator, all the ellipses around the entity are the attributes of those entities, at the second stage those entities will be the tables of the data base and the attributes are the fields of the tables.
For example to trace that ER diagram, the administrator specifies categories of the shops, while shops belong to categories.
Recommendations written by customers, recommendations related to certain product.
Shop's management add items, and add products, also they introduce offers for sales or whatever. In that way we trace that ER – Diagram.

### 3-3 Online Shopping Steps

Buying process is made up of several stages, site visitors could fall into any one of those steps.

Step 1: Select products:

Selecting products can be done by login to the mall site, then select a store follow by choosing an item from the store and purchase the item

Step 2: Confirmation

A confirmation page will be presented with the total value of the orders of the products want to be purchased. The information of the buyer is required to be entering such as the name, email address and postal address.

Step 3: Credit card details

The credit card processor's site will be presented with credit card details (the name, number and expiry date on the credit card). This will be done through a secure connection with state-of-the-art encryption. The transaction will be approved through process subject to the verification of credit card number and other required security issue.

Step 4: Email receipt and installation codes

Right after the transaction approved, automatically email sent to Customer Service with all the product installation codes. An email will be received by customer right after the transaction approved with an electronic receipt of the order. This normally takes a few seconds,

Step 5: Product delivery

The product will be delivered to the address given earlier in customer registration. Online credit card processing is the foundation of e-commerce

### 3-4 Online Payment Transaction

Online transaction has really brought payment easier than ever. Now, wherever you are in the world, you can get your

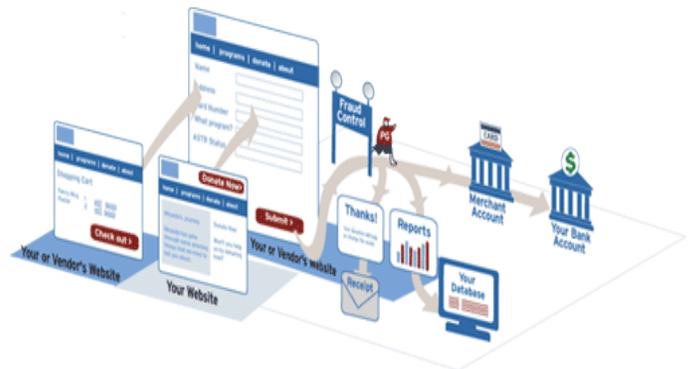

payments processed. The process works, in 8 simple steps:

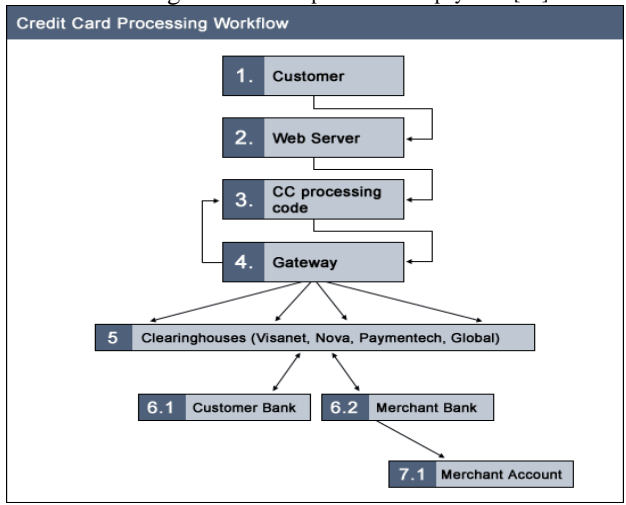

Fig. 3: The whole process of the payment [30]

Fig. 4: Credit Card Processing Workflow [29]

1- The process begins with a customer as in fig (4) box 1 who wants a purchase of a certain product. Most of the time, many online business and product sellers, already have a software application which shows all the products or servicing they are rendering. The customer just needs to navigate and click the product of his choice.
2- **T**he customer is online, typically looking at an HTML form. This form collects the customer's credit card information and sends it to the server for processing. The user fills out the form and then clicks Submit.

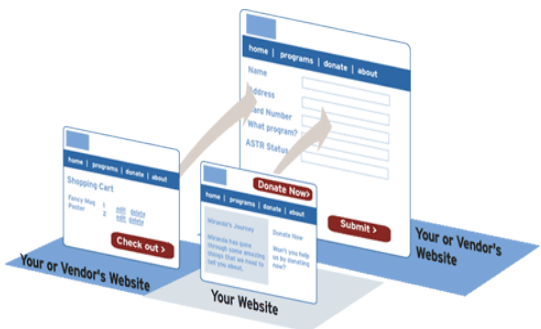

Fig. 3-1: Entering Customer Information

3- The server receives the information POSTed in the form the user submitted. The server then sends the information to code that resides on the server for processing.

4- The processing code receives the information from the Web server and validates the data entered by the user. If the data is valid, the code formats the data into a format that the gateway can understand. The code then sends the formatted data to the gateway. In effect, the code is asking the gateway whether the credit card is a good card and whether it can do the transaction.

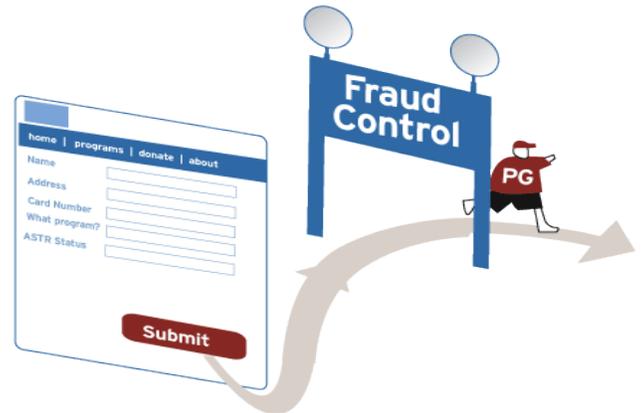

Fig. 3-2: Validation Process

5- The gateway validates the card, and checks the amount weather is available or not in the users account. If the card is good and the funds are available, the gateway sends an approved message back to the code (box 3); if not, the gateway sends a declined message back to the code. For providing this service, the gateway charges the merchant money.
6- Then at the gateway the transactions are batched through to the clearinghouse which is a financial institution stands between two clearing firms Box 5. The clearinghouses receive transactions from all the gateways, and transfer the monies from bank to bank.

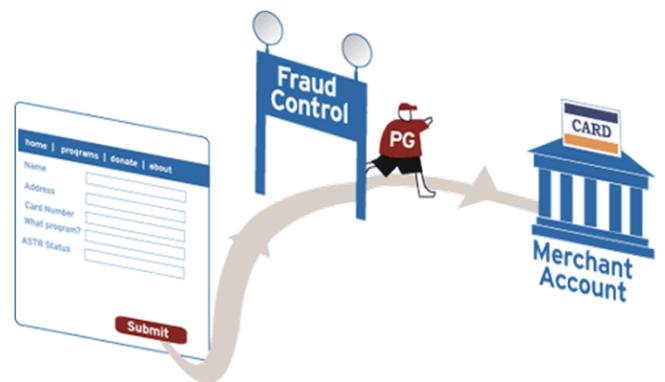

Fig. 3-3: batching the transactions

7- As the clearinghouses batch the transactions they receive, they transfer money from the customer's bank (6.1) to the merchant's bank (6.2).

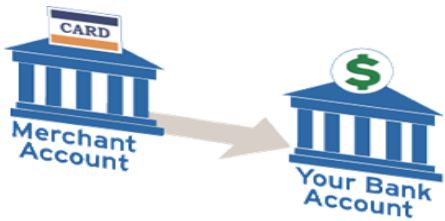

Fig. 3-4: Transfer the money from customer's account to merchant's account

8- The merchant's bank receives the transactions from a clearinghouse and then transfers the appropriate amount of money for the customer transaction (started in box 1) into the Merchant's Card Not Present merchant account (7). For providing the Merchant account, the bank will charge various fees. Different banks have very different fee structures; contact your bank for details on Card Not Present merchant account costs.

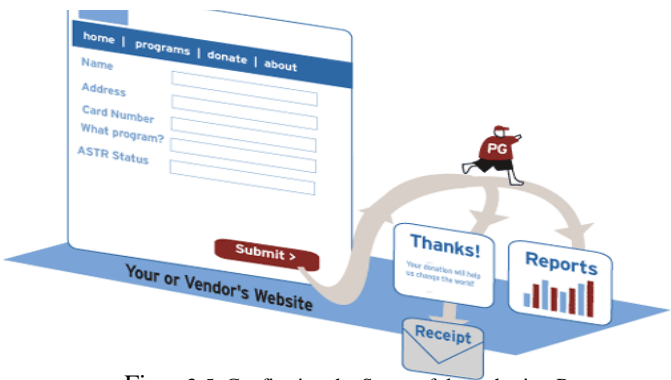

Fig. 3-5: Confirming the Successful purchasing Process

The customer will receive a receipt for the purchase items by email confirming that the payment went through and the transaction is successfully done.

## 4- CONCLUSION AND FUTURE WORK

3D virtual environment online present a three-dimensional user interface that are specialized to navigate and manipulates physical objects inside the virtual world. So that will make the customers think they are physically there. This technology offers the ability to simulate in the real word environments and gives the customer a flexibility to engage with the activities that are closer to experiencing the products and services. This technique is a new attractive environment led to enhance customer's beliefs, attitudes, and behaviors toward the products and provide the user with convenience and entertainment. This environment show great potential for enhancing e-commerce systems and provide customers information about a product and real shopping environment.

Concerning the future work, this research will continue to expand the idea of 3D virtual environment as applications on an Android cell phone.